\documentclass{iopart}

\usepackage{graphicx}
\usepackage{bm}
\usepackage[latin1]{inputenc}
\usepackage{color}
\usepackage{amssymb}
\usepackage{times}
\usepackage{colordvi}
\usepackage{color}
\usepackage{epsfig}

\begin{document}

\title{Long-term stable squeezed vacuum state of light for gravitational wave detectors}

\author{Alexander Khalaidovski, Henning Vahlbruch, Nico Lastzka, Christian Gr\"af, Karsten Danzmann, Hartmut Grote and Roman Schnabel}
\address{Max-Planck-Institut f\"ur Gravitationsphysik (Albert-Einstein-Institut) and\\
\mbox{Institut f\"ur Gravitationsphysik, Leibniz Universit\"at Hannover}\\
Callinstr.~38, 30167 Hannover, Germany}
\ead{Roman.Schnabel@aei.mpg.de}

\date{\today}

\begin{abstract}
Currently, the German/British gravitational wave detector GEO\,600 is being upgraded in course of the \emph{GEO-HF} program. One part of this upgrade consists of the integration of a squeezed light laser to nonclassically improve the detection sensitivity at frequencies where the instrument is limited by shot noise. This has been achieved recently~\cite{SqzNatPhys}. The permanent employment of squeezed light in gravitational wave observatories requires a long-term stability of the generated squeezed state. In this paper, we discuss an unwanted mechanism that can lead to a varying squeezing factor along with a changing phase of the squeezed field. We present an extension of the implemented coherent control scheme~\cite{{VCHetal06}} that allowed an increase in the long-term stability of the GEO\,600 squeezed light laser. With it, a quantum noise reduction by more than 9\,dB in the frequency band of 10\,Hz\,--\,10\,kHz was observed over up to 20 hours with a duty cycle of more than 99\,\%.\\

\end{abstract}
\maketitle

\section{Introduction}
\indent In the last decades, a network of kilometre-scale Michelson-type laser interferometers has been set up aiming at the first direct detection of gravitational waves (GWs)~\cite{A09,A08,G10}. These earth-bound GW detectors target signals at audio frequencies in a band of ca.~10\,Hz\,--\,10\,kHz. While at lower and intermediate frequencies the measurement sensitivity is, at the moment, still limited by seismic noise (below 50 Hz) and thermal or technical noise (several hundred Hertz), at higher frequencies only the quantum nature of light inhibits a more sensitive measurement. Currently, the observatories undergo an update program, bringing them to the second or \emph{Advanced} detector generation with a design sensitivity limited by shot noise at lower frequencies (approximately 300\,--\,400\,Hz in case of Advanced LIGO~\cite{H09}). This quantum noise arises from the zero-point fluctuations of the electro-magnetic field being in it's ground (so-called \emph{vacuum}) state~\cite{SMML10}. A possibility to improve the signal-to-noise ratio is the employment of \emph{squeezed states of light}~\cite{Y76, GK04}, as was first proposed by Caves in~\cite{C81}. An overwiev of the field of study can be found in the review article~\cite{SMML10}.\\
\indent The first experimental demonstration of squeezed states of light was achieved in the 1980s by Slusher \emph{et al.}~\cite{SHYMV85}, closely followed by other research teams \cite{WXK87, PXKH88}. In the following two decades, interest moved away from proof of principle experiments to the construction of robust sources of squeezing with the main concern of improving the squeezing figures. For this purpose, below-threshold optical parametric oscillators proved to be very efficient squeezed light sources. In these, squeezing is generated via parametrical down-conversion in a $\chi^{(2)}$ nonlinear crystal. Strong squeezing at the carrier wavelength of the currently operated GW detectors, being 1064\,nm, can e.g.~be generated in MgO:LiNbO$_3$~\cite{VMCHFLGDS08, MVLDS10} or periodically poled KTP (PPKTP)~\cite{ESBHVMMS10}, pumped by a second-harmonic 532\,nm beam. Additional requirements on a squeezed light source for GW detectors were imposed by their audio-frequency operation band. Proof of principle experiments therefore included the development of novel control schemes to generate squeezing at audio frequencies \cite{VCHetal06, MGBetal04, MMGetal05, VCHetal07} as well as the demonstration of high degrees of squeezing~\cite{MVLDS10, ESBHVMMS10}. The implementation of squeezed light was tested in a suspended GW prototype detector \cite{GMMetal08}. Thereafter, a squeezed light laser was constructed for the first upgrade of a large-scale GW detector, the German-British collaboration project GEO\,600, and first characterization results published in~\cite{VKLGDS10}. Recently, a nonclassically improved detection sensitivity of the GW detector GEO\,600 by up to 3.4\,dB was reported in~\cite{SqzNatPhys}.\\
\indent The observation of a squeezing spectrum at low audio frequencies (e.g.~down to 1\,Hz as reported in Ref.~\cite{VCHetal07}) requires the experimental setups involved to be stable at the timescale of several tens of minutes. This was achieved realizing appropriate control loops to stabilize the length of the optical cavities with respect to the laser field, the temperature of the nonlinear crystals as well as the phase relation of the generated squeezed field with respect to the pump field and to a local oscillator field employed for squeezing characterization. A permanent employment in a GW observatory, however, imposed a furthermore increased demand on the GEO\,600 squeezed light laser: a (preferably high) squeezing factor needs to be continuously provided on timescales of hours to days (in the following referred to as \emph{long-term stable}) in order to ensure a non-varying observatory sensitivity. Therefore, beyond the proof of principle experiments, the GEO\,600 squeezed light laser setup was designed to have a high mechanical stability and was furthermore extended by a long-range PZT actuator for the control of the optical phase of the squeezed field. During the characterization of the long-term performance of the system we found, however, the stability of the squeezing degree to depend on the power stability of the 532\,nm pump beam for the coherent control scheme employed. In the following, we summarize the basic principle of the initially implemented coherent control scheme and discuss the effects of a time-dependent pump power value on the squeezing degree. Finally, the extension of the coherent control scheme by a pump power stabilization scheme is discussed. This extension allowed to observe on average 9\,dB of squeezing at audio frequencies over a time period of 20 hours with a duty cycle exceeding 99\,\%.

\section{Experimental setup}
\begin{figure}[t]
\begin{center}
		\includegraphics[width=\linewidth]{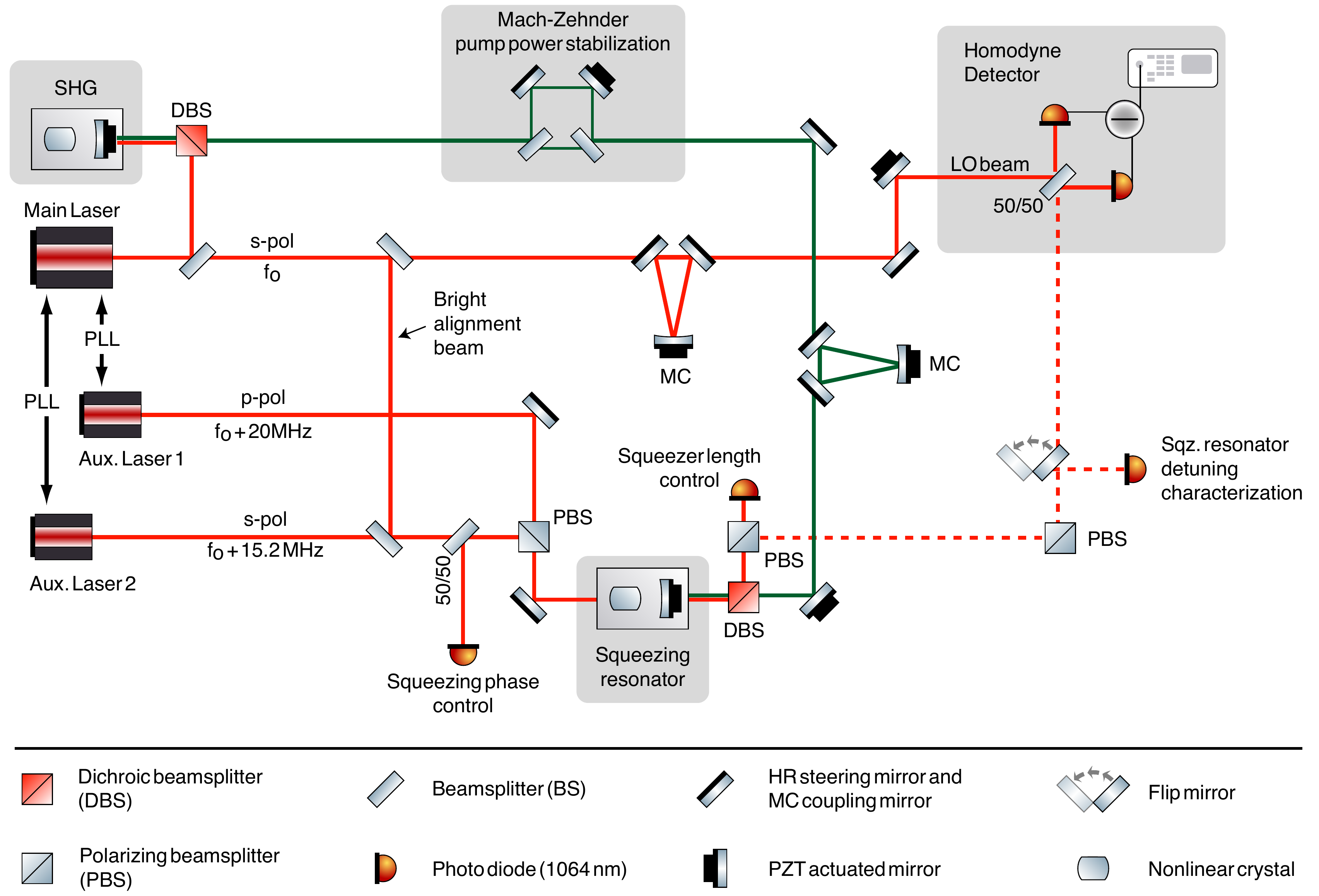}
\end{center}
		\caption{Simplified sketch of the optical layout for the long-term stable generation of squeezing at audio frequencies. Auxiliary laser 1 is used to stabilize the length of the squeezing resonator and auxiliary laser 2 allows controlling the phase of the squeezed field. Both auxiliary lasers are phase locked (PLL) to the main 2\,W laser (with f$_0$ denominating the optical carrier frequency). We found the optical length of the squeezing resonator to depend on the 532\,nm pump power value. For characterization, the s-polarized bright alignment beam was employed. Detected in transmittance of the squeezing resonator, it directly allows mapping out the cavity's resonant curve. The optical gain was close to unity for the measurement.	The implemented pump power stabilization was realized via a Mach-Zehnder (MZ) type interferometer.}
\label{figure1}
\end{figure}
\indent A simplified sketch of the squeezed light laser setup is presented in Fig.~\ref{figure1}. A detailed description of the initial optical layout (without the pump power stabilization) as well as of the control scheme can be found in~\cite{VKLGDS10}. To maintain clarity, only the most important optical components are shown. The experiment is driven by a monolithic non-planar Nd:YAG ring laser (NPRO) of 2\,W single-mode output power at 1064\,nm. One part of this beam is frequency up-converted in a second-harmonic generator (SHG), which uses 7\% MgO doped LiNbO$_3$ as nonlinear $\chi^{(2)}$ medium. This 532\,nm \emph{pump beam} is filtered by a mode-cleaning travelling-wave resonator to attenuate high-frequency phase noise which has been shown to diminish the maximal degree of squeezing achievable~\cite{TYYF07, FHDFS06}. Subsequently, the pump beam is injected into the squeezed light source which consists of a periodically poled potassium titanyl phosphate (PPKTP) crystal placed in a standing-wave hemilithic cavity.\\
\indent The long-term stable generation of squeezed states at audio frequencies requires the implementation of a control scheme that avoids the introduction of technical laser noise to the squeezed vacuum state. This control scheme is realized using two additional NPRO laser sources. Both lasers are phase locked to the main 2\,W laser (which in turn is phase locked to the main GEO\,600 laser when operating the setup at the detector site). The first auxiliary laser beam is used to control the length of the squeezing resonator via a Pound-Drever-Hall (PDH) locking scheme~\cite{DHKHFMW83}. This field has a polarization orthogonal to the one of the squeezed field. Due to the crystal birefringence, the frequency of the length control beam has to be shifted by 20\,MHz to fulfill the same resonance condition on phase-matching temperature as the squeezed field. Thereby, no technical laser noise is introduced into the squeezed field at the fundamental frequency.\\
\indent The second auxiliary laser allows controlling the orientation of the squeezing ellipse with respect to the pump field. This \textit{coherent control field (CCF)} has the same polarization as the squeezed beam and is shifted by a frequency offset of $f_{\rm{CCF}}\,=\,$15.2\,MHz. To stabilize the relative phase between the control field and the pump field, an error signal is generated by detecting the control beam back-reflected from the squeezing resonator and by demodulating the photocurrent measured at twice the offset frequency~\cite{CVDS07}.\\
\indent The generated squeezing is measured in an on-board diagnostic balanced homodyne detector. For this purpose, a small fraction of the main 1064\,nm beam is used as a local oscillator (LO) beam, which is also filtered by a ring mode-cleaner. Another loop employing the coherent control field transmitted by the squeezing resonator allows the control of the phase relation between the generated squeezed field and the LO of the homodyne detector.\\
\indent To enable permanent operations, the experiment was interfaced with a real-time UNIX based control system. It allows the monitoring of all relevant experimental channels using AD-converters and also a remote control of the entire experiment.

\section{Influence of pump power fluctuations}
\indent The squeezing output in several ways depends on the 532\,nm pump beam power. Pump power fluctuations will change the degree of squeezing and anti-squeezing, as well as the angle of the squeezing ellipse, as shown below. One reason for the varying pump power is that the 1064\,nm laser beam used to pump the SHG device is not stabilized by any means. The fluctuations of the 1064\,nm laser power therefore directly couple into the 532\,nm power value. Please note that hereby the nonlinear coupling mechanism (due to the power dependence of the SHG conversion efficiency) may result in an even greater relative power change for the second harmonic beam. Furthermore, degradation of beam alignment into the SHG (being a weeks- or month-timescale effect) will also lead to a slowly changing pump power level. This time-dependence of the 532\,nm pump power is not compensated by the control loops described in~\cite{VKLGDS10}. A first characterization of the degree of squeezing has shown the latter to slowly degrade without a pump power stabilization employed. Though a detailed analysis of the power fluctuation has not been performed, values of several percent have been observed.\\
\indent The reason for the effect lies, as we assume, in the non-vanishing absorption coefficient of the nonlinear crystal for 532\,nm light (the reported absorption coefficients cover a range of 0.5\,\%/cm - 4.5\,\%/cm~\cite{PhDMeier}). A pump power change results in a change of the absorbed power and thus in a change of the crystal temperature. The changed temperature gradient inside the crystal is not sensed in an efficient way by the implemented temperature stabilization loop and therefore cannot be compensated for. Please note that this problem is not fundamental and in principle can be solved by a calibrated temperature stabilization scheme. The alternative pump power stabilization solution presented in Chapter~\ref{sec.discussion} is, however, both, more simple and precise. Due to the temperature dependence of the refractive index, a pump power variation results in a change of the optical cavity length.\\
\indent To evaluate the expected effect for the squeezed light laser, let us assume the following starting condition: The length of the squeezing resonator is stabilized using the p-polarized control beam and the degeneracy condition between the s- and the frequency-shifted p-polarized light is adjusted using the crystal temperature as a fine-adjustment tool. When operated in this way, the temperature is still optimized in regard to the phase-matching condition and the squeezing resonator delivers the maximal amount of squeezing possible at the chosen pump power. Now the pump power increases and the increased thermal load leads to a cavity length increase. While the resonator is still held on resonance with the PDH control loop, the polarization degeneracy is no longer fulfilled to 100\,\%. Thus, the increased power results in a cavity that is perfectly resonant for p-polarized light but slightly detuned from resonance for the s-polarized squeezed field.\\
\indent To obtain quantitative results for the detuning of the squeezing resonator to be expected at a certain pump power deviation from the set value, 
\begin{figure}[t]
\begin{center}
		\includegraphics[width=\linewidth]{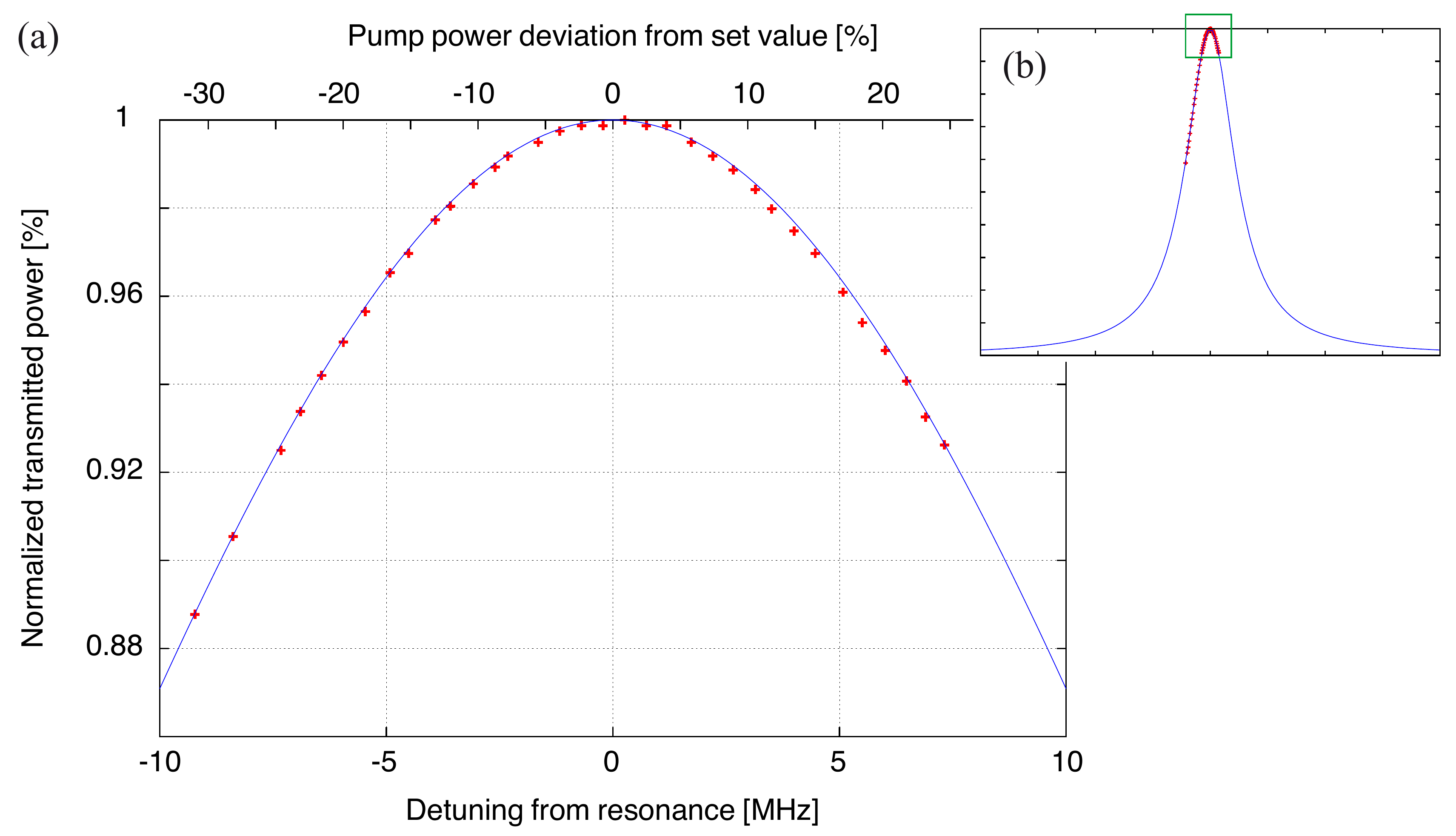}
\end{center}
		\caption{(a) Measurement of the alignment beam power transmitted through the squeezing resonator as a function of pump power. A power of 34.5\,mW, corresponding to the envisioned GEO\,600 operations condition, was employed as set value. The measured data (red crosses) are in good agreement with the simulation of the cavity resonant curve (blue line) employing experimental parameters. (b) Frequency zoom-out showing the location of the measured data with respect to the resonant curve.}
\label{figure2}
\end{figure}
the following characterization measurement was performed. For this purpose, a beam tapped off from the main $f_0$ laser beam was employed (\textit{bright alignment beam} in Fig.~\ref{figure1}), having the same frequency and polarization properties as the squeezed field. The length of the squeezing resonator was stabilized using the p-polarized field, and a pump power of $P_{\rm{532\,nm}}\,=\,$34.5\,mW was injected into the squeezing resonator, while the coherent control field was blocked. The crystal temperature was optimized for a polarization degeneration and the pump beam polarization shifted by 90°. This ensured that lowest nonlinear conversion took place while the 532\,nm power was still present and absorbed inside the crystal. Now the pump power was varied and the change in the alignment beam power (being proportional to the intracavity power and hence to the cavity's resonant curve) recorded by a diagnostic photo detector (addressed as \textit{sqz.~resonator detuning characterization} in Fig.~\ref{figure1}). This measurement directly allowed deducing the resonator detuning as a function of the relative pump power change. The normalized results are shown in Fig.~\ref{figure2}. The measured data (red crosses) are in good agreement with the simulation of the cavity resonant curve (blue line).\\
\begin{figure}[b]
\begin{center}
		\includegraphics[width=\linewidth]{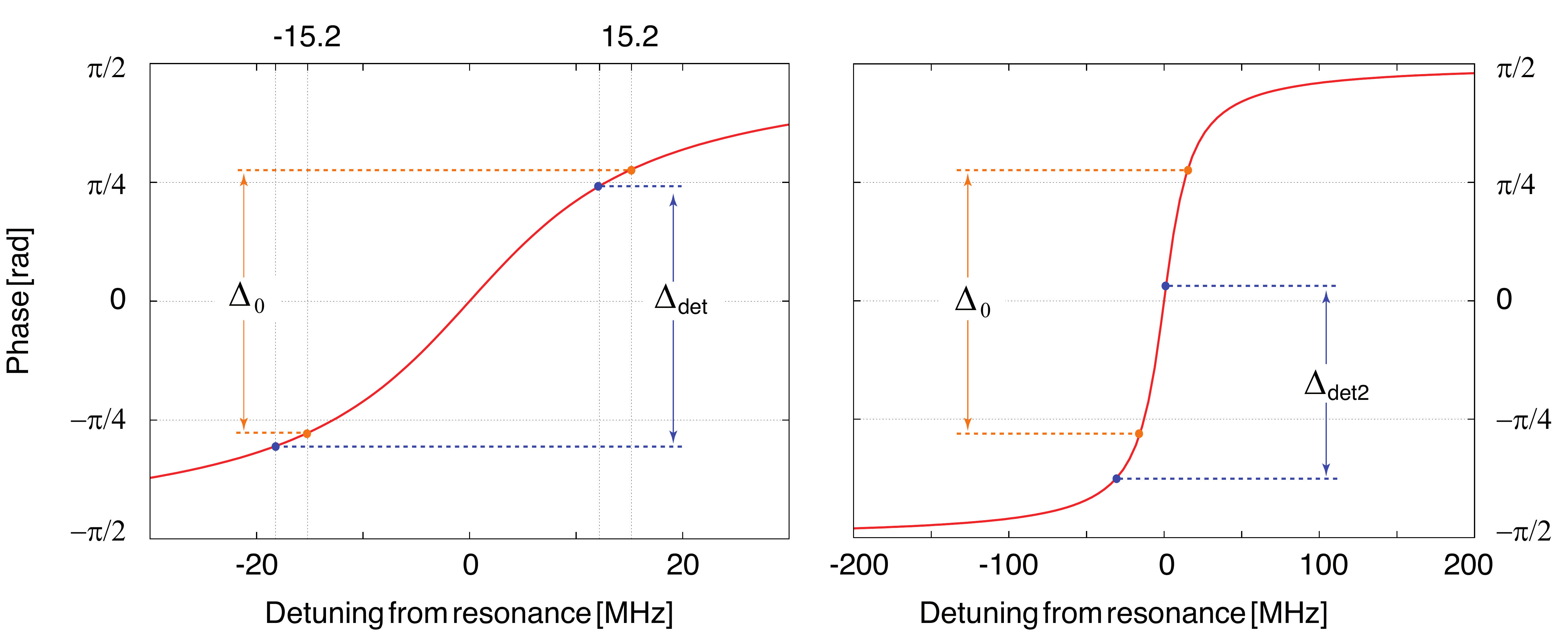}
\end{center}
		\caption{Left: Phase rotation experienced by an optical field transmitted through the squeezing resonator in dependence of its detuning. The phase difference between the two CCF sidebands, serving as reference for the squeezed field phase stabilization, is always larger for a resonant ($\Delta_0$) than for a detuned ($\Delta_{\rm{det}}$) cavity. The difference $\Delta_0 - \Delta_{\rm{det}}$ is small for detunings much smaller than the cavity linewidth and reaches a constant maximal value for frequencies much larger than the lindewidth. The shown frequency offset was chosen to be 3\,MHz, corresponding to the effect of a 10\,\% variation of the 532\,nm pump power. Right: Zoom out at large detuning values for a visual illustration of the effect described with an arbitrary example frequency offset of 15\,MHz.}
\label{figure3}
\end{figure}
\indent When the squeezing resonator is detuned from resonance, the generated squeezed field is effected in several ways. First, the coherent control scheme is based on a single sideband field used for stabilization of the squeezing ellipse angle. The other sideband at the frequency $f_0 - f_{\rm{CCF}}$ is generated by the nonlinear $\chi^{(2)}$ interaction. The phase relation of both sidebands is then used as reference in the control scheme. To provide an intuitive understanding of the detuning effect, Fig.~\ref{figure3} shows the phase shift
\begin{equation}
\Phi(f) = {\rm{arg}}\left(\frac{\tau_1\tau_2 \exp{\frac{{\rm{i}} l 2\pi f}{c}}}{1-\rho_1\rho_2 \exp{\frac{2 {\rm{i}} l 2\pi f}{c}}}\right)
\label{equation1}
\end{equation}
experienced by an optical field transmitted through the squeezing resonator. Hereby, $\tau_{1/2}$ and $\rho_{1/2}$ denote the amplitude transmittances and reflectivities, respectively, of the resonator coupling mirrors. In the coherent control scheme, the optical phase difference $\Delta_0 = \Phi(f_{\rm{CCF}})-\Phi(-f_{\rm{CCF}})$ of the two sidebands is used as reference to stabilize the phase of the squeezed field. When the cavity is now detuned from resonance by $f_{\rm{det}}$, both sidebands experience a phase rotation different from the previous condition as illustrated in Fig.~\ref{figure3}. As a consequence, the squeezing ellipse experiences a rotation by an angle depending on the detuning and hence on the pump power variation. This rotation directly translates into the degree of squeezing detected at the diagnostic homodyne detector as
\begin{equation}
V'_{{\rm{a/s}}, \Theta} = V_{\rm{a/s}}\,\cos^2{\Theta} 
+ V_{\rm{s/a}}\,\sin^2{\Theta}.
\label{eq.homodegrading}
\end{equation}
$V_{\rm{s/a}}$ denominates the value of squeezing and anti-squeezing, respectively, while $\Theta$ is the rotation angle. Trace (a) of Fig~\ref{figure4} shows the simulated degradation of squeezing under a fixed readout angle ($V'_{\rm{a/s}, \Theta}$) that was initially optimized for the highest squeezing at the nominal pump power. For the simulation, a state with 9.3\,dB of squeezing and 16.75\,dB of anti-squeezing was assumed.\\
\begin{figure}[t]
\begin{center}
		\includegraphics[width=0.8\linewidth]{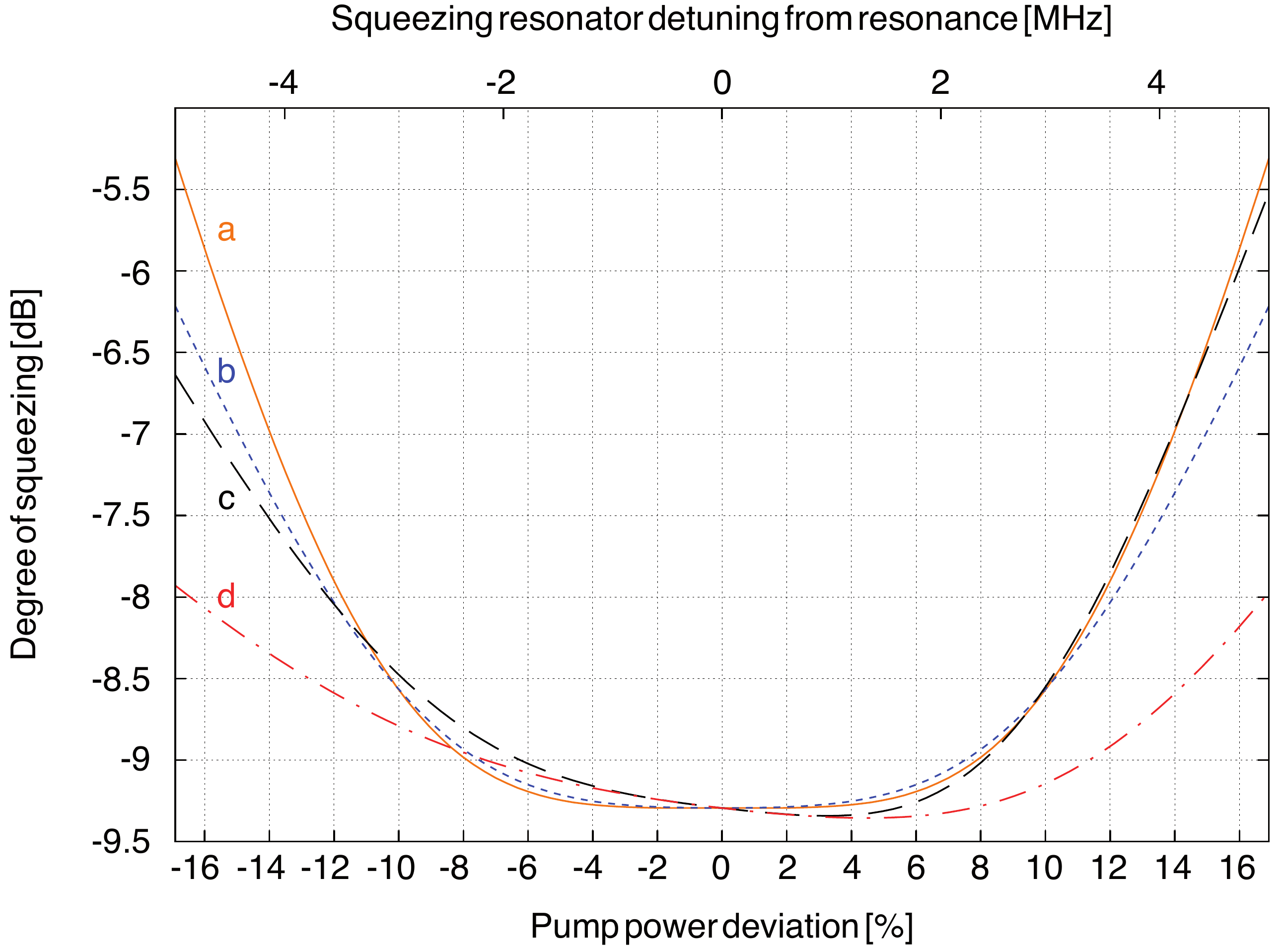}
\end{center}
		\caption{Simulated degree of squeezing detected at a fixed homodyne angle, initially optimized for the nominal pump power of 34.5\,mW, as function of pump power. The upper x-axis additionally shows the corresponding detuning of the squeezing resonator (for s-polarized light). Trace	(a) considers the rotation of the squeezing ellipse directly induced by the detuning.	(b) takes into account the reduced squeezing output of a detuned resonator. (c) additionally accounts for the optical gain changing with pump power. (d) finally takes into account the detuning-induced rotation of the homodyne readout quadrature. Please note that this simulation does not account for the effect of the detuning-dependent nonlinear gain on the CCF sidebands or on the error signal magnitude.}
\label{figure4}
\end{figure}
\indent The homodyne angle stabilization loop is effected by the resonator detuning in the same way. In this case, however, the optical beat of the double-sideband CCF field and the LO field at frequency $f_0$ is employed for stabilization. It is hence the average value of the phase differences between the $f_0$ field and the two sideband fields that serves as reference. Thus, the readout quadrature is co-rotating with the squeezed field quadrature and therefore to some extent compensates for the squeezing degradation as shown in trace (d) of Fig.~\ref{figure4}.\\
\indent An additional parameter is the detuning-dependent nonlinear gain experienced by both sidebands which will result in a change of the sideband amplitudes and hence in a further effect on the resulting phase difference. Please note that this effect is not accounted for by the simulation results shown in Fig.~\ref{figure4}.\\
\indent The anti-squeezing and squeezing variances
\begin{equation}
R_{\rm{a/s}} = 1\pm \eta_{\rm{tot}} \frac{4\sqrt{P/P_{\rm{th}}}}{(1 \mp \sqrt{P/P_{\rm{th}}})^2+4 \kappa^2}
\label{eq.mz1}
\end{equation}
directly depend on the pump power $P$~\cite{TYYF07}. In this expression, \textit{a} and \textit{s} denote the anti-squeezing and the squeezing, respectively, $\eta$ describes the total detection efficiency and $P_{\rm{th}}$ is the OPO threshold power. The normalized frequency
\begin{equation}
\kappa = \frac{2\pi f}{\gamma} \qquad\qquad {\rm{with}} \qquad\qquad \gamma = \frac{c(T+L)}{l}
\label{eq.mz2}
\end{equation}
is a function of cavity parameters, where $T$ is the transmittance of the output coupling mirror, $L$ is the intra-cavity loss, $l$ the optical cavity round-trip length and $f$ is the frequency. Please note that the coupling of the anti-squeezing into the squeezing introduced by the ellipse rotation may effectively result in a reduction of the observed squeezing degree for increasing pump powers and vice versa as shown in traces (b) and (c) of Fig.~\ref{figure4}.\\
\indent The magnitude of the error signal used for the homodyne angle stabilization loop depends on the intensity of the coherent control beam transmitted through the squeezing resonator as well as on its nonlinear gain. Both values will change when a detuning is introduced. In case the zero-crossing point of the error signal is chosen as operation point, an intensity change does not have an effect on the detection angle. Since in the homodyne locking scheme an electronic offset is added to the error signal in order to adjust the desired ellipse orientation, an operation at the mid-point of the sinusoidal error signal will usually not be the case. Depending on the operation point, a changing error signal magnitude will result in a rotation of the readout quadrature, which is the stronger the closer the operation point is to the turning point of the error signal.\\
\indent As long as the cavity is held on resonance, the phase shift experienced by the squeezed field is equal in magnitude (with an opposite sign) for the sidebands at the upper and the lower Fourier frequencies. This results in a frequency-independent orientation of the squeezing ellipse in phase space. When the resonance condition is no longer fulfilled, the phase shifts experienced by the upper and the lower sidebands of the squeezed field are no longer equal, which leads to a frequency-dependent rotation of the squeezing ellipse, as was shown in~\cite{CVHFLDS05}. In the discussed case, this effect is, however, negligible in a first-order consideration since frequencies far smaller than the cavity linewidth are considered.\\
\indent Summarizing, pump power fluctuations result in an alteration of the squeezing strength as well as in a rotation of the squeezing ellipse. While the first effect is permanent, the second one can be compensated for by manipulating either the squeezing phase or the homodyne readout angle. An experiment relying on a long-term stability of the squeezing angle will, however, suffer severe degradation of the squeezing strength detected by the homodyne detector at one fixed readout angle. For typical SHG performances, a pump power stabilization scheme seems hence mandatory for any long-term operations employing our coherent control topology.\\ 

\section{Long-term stable squeezing and discussion}
\label{sec.discussion}
\begin{figure}[t]
\begin{center}
		\includegraphics[width=\linewidth]{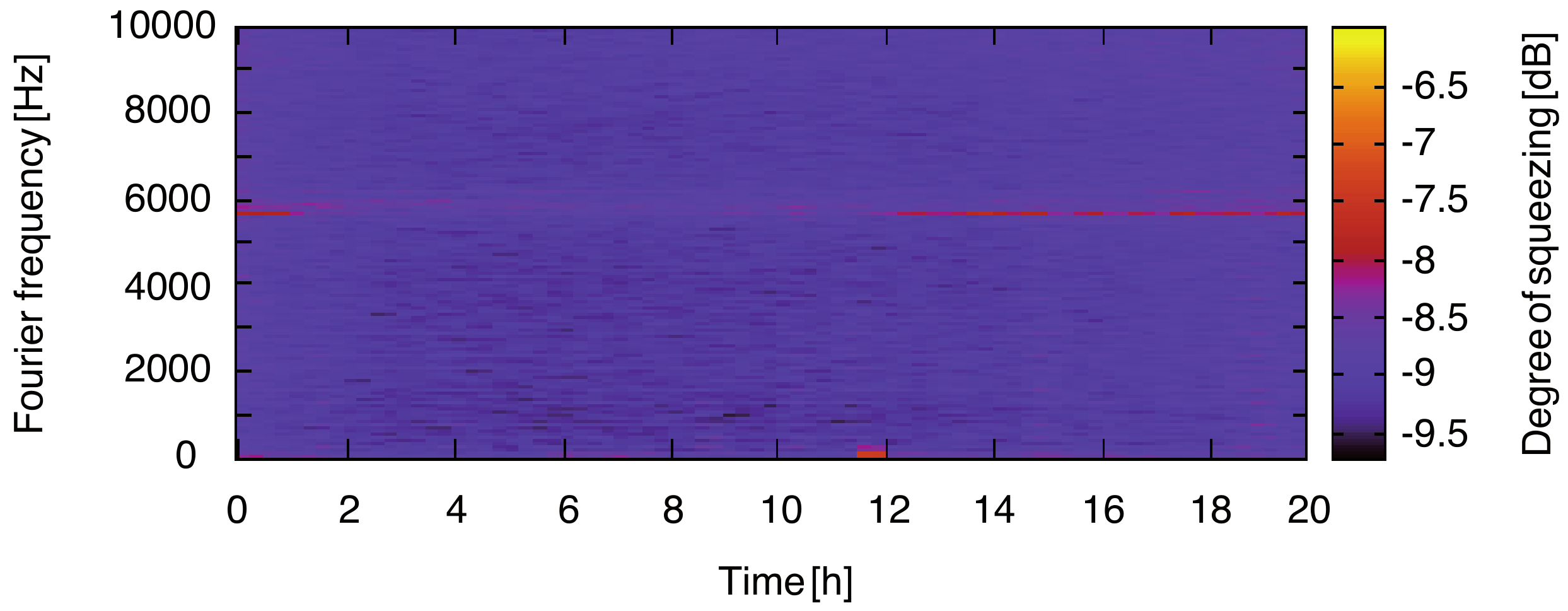}
\end{center}
\caption{Squeezing spectrogram. The time resolution is 15 minutes, each time bin was generated by the application of a FFT to a 60 sec long dataset. The data was recorded continuously at the output of the homodyne detector through a fast CDS channel. The duty cycle of the squeezed light laser for this long-term characterization was higher than 99$\%$, 8 lock-losses were observed over 20 hours. The longest continuous lock extended over 8.5 hours. The lower squeezing value around 6\,kHz originates from electronic pick-up at the GEO\,600 detector site.}
\label{figure5}
\end{figure}
The implemented pump power stabilization is, as shown in Fig.~\ref{figure1}, realized via a Mach-Zehnder type interferometer. One of the steering mirrors is mounted on a PZT actuator, thereby allowing to change the optical power transmitted by the interferometer. For the stabilization scheme, a small fraction of the 532\,nm pump beam is tapped off close to the 
squeezed light source and monitored with a photo detector. The unity gain frequency was 1\,kHz.\\
\indent Fig.~\ref{figure5} shows the long-term behaviour of the squeezed light source observed during characterization at the GEO\,600 detector site. For this measurement, the main squeezing laser was phase-locked to the GEO\,600 laser. The squeezed light source was operated with a 532\,nm pump power of 35\,mW, resulting in a measured averaged squeezing value of about 9\,dB. Such operation conditions correspond to more than 10\,dB of squeezing exiting the squeezed light laser breadboard and directed to GEO\,600. The longest quasi-continuous squeezing observation extended over 20 hours, the test run was terminated by maintenance work at GEO\,600. The longest total lock of the entire system extended over 8.5 hours, the overall duty cycle exceeded the value of 99\,\%.\\
\indent When GEO\,600 is operated in science mode, the duty cycle is in the order of 90\,\%~\cite{G10} with typical lock durations extending from several hours up to 40\,--\,60 hours, being larger than for the squeezed light laser. This difference should, however, not significantly affect the duty cycle of a squeezed-input GEO\,600. In case of lock-loss, a shutter automatically blocks the squeezing beam input path, thereby ensuring that the detector sensitivity is not degraded due to anti-squeezing input. Hence, the detection sensitivity will be improved by squeezing most of the time and be equal to the classical sensitivity when the squeezed light laser is unlocked. A lock-loss of GEO\,600, on the other hand, may also lead to a lock-loss of the squeezed light source. Because the lock acqusition time of the latter is with max.\ 15 seconds much shorter than the approximately 10 minutes of GEO\,600, no additional performance degradation is expected.\\
\indent Due to the GEO\,-\,HF update program and to the works conducted at the detector site, no undisturbed extended long-term characterization of the system could be performed. A permanent observatory operation with squeezing input after the end of the GEO\,-\,HF upgrade will allow deducing the stability of the generated squeezing degree on a timescale of weeks or months. The results reported in this paper, however, let us expect that the performance of the squeezed light laser will not significantly affect the duty cycle of GEO\,600, while more than 10\,dB of squeezing will be injected over most of the time.

\section{Summary}
A permanent employment of squeezed light in GW observatories imposes the requirement of a multi-hours-timescale stability along with an automatized control with fast lock re-acquisition times on the squeezed light laser. In the coherent control scheme used to generate squeezed light over the entire earth-bound GW detector frequency band of 10\,Hz\,--\,10\,kHz, the stability of the squeezing degree, besides a multiplicity of parameters addressed in previous experiments, for long observation times directly depends on the power stability of the 532 nm pump beam. In this paper, we discussed how a drifting pump beam power translates into a degrading degree of squeezing along with a varying phase of the squeezed field. Finally, we presented an extension of the coherent control scheme which allowed to observe about 9\,dB of squeezing over up to 20 hours. The longest continuous lock extended over 8.5 hours. After a lock-loss, the squeezing output is restored after no longer than 15 seconds. To the best of our knowledge, this constitutes the highest degree of squeezing ever reported at audio frequencies, along with the longest continuous operation of a squeezed light source ever achieved in this frequency band. The duty cycle of the squeezed light laser exceeded 99\,\% and thereby does not constitute a limiting factor for the duty cycle of a gravitational wave interferometer.\\

\ack
We would like to thank Albrecht R\"udiger and Aiko Samblowski for helpful and valuable discussions. This work has been supported by the international Max Planck Research School (IMPRS) and the cluster of excellence QUEST (Centre for Quantum Engineering and Space-Time Research).

\end{document}